\documentclass[aps,prd]{revtex4}

\usepackage{amsmath}
\usepackage{amssymb}
\usepackage{bm}
\usepackage{graphicx}
\usepackage{multirow}
\usepackage{color,soul}

\usepackage{textcomp}

\allowdisplaybreaks[1]

\begin{document}

\title{LHC di-dijet excesses as signals of fourth-generation tetraquarks}

\author{Hsiang-nan Li}
\affiliation{Institute of Physics, Academia Sinica,
Taipei, Taiwan 115, Republic of China}

\date{\today}

\begin{abstract}

We postulate that the excesses of di-dijet events observed at the LHC are attributed to the
production of four fourth-generation quarks $b'$ with a mass $m_{b'}\approx 2$ TeV at
few-TeV scales. The di-dijet signals around the four-jet invariant mass $m_{4j}\approx 8$ 
TeV arise from a resonant $b'b'\bar b'\bar b'$ tetraquark production, where the dijet 
resonances of masses about 2 TeV correspond to $b'\bar b'$ first excited states 
(color-octet scalars with the principal quantum number $n=2$) in a Yukawa potential created 
by Higgs boson exchanges. Those around $m_{4j}\approx 3.6$ TeV originate from a 
non-resonant $b'b'\bar b'\bar b'$ production, where the dijet resonances of masses 0.95 TeV 
correspond to $b'\bar b'$ ground states (color-octet vectors with $n=1$). It is shown 
that a $b'\bar b'$ system with $m_{b'}\approx 2$ TeV in the Yukawa potential does generate 
the aforementioned bound state spectrum. We then illustrate that the observed excesses can 
be accommodated in our setup by translating the fourth-generation model to the effective 
theories containing color-octet scalars and vectors available in the literature. 
Simply speaking, our scenario can be viewed as a TeV-scale 
analogue of the search for a fully charmed tetraquark via the four-muon channels 
$X(6900)\to (c\bar c)(c\bar c)\to 4\mu$ at a GeV scale.

\end{abstract}

\maketitle

\section{INTRODUCTION}
\label{sec1}

The resonant production of pairs of dijet resonances with the same invariant mass has 
been searched for by the ATLAS \cite{ATLAS:2023ssk} and CMS \cite{CMS:2020zti} 
Collaborations. The process proceeds through $pp \to Y \to XX \to (jj)(jj)$, 
where the intermediate state $Y$ decays into the identical dijet resonances $X$, for which 
both jets $j$ are individually reconstructed. The particle $Y$ may form a broad resonance, 
whereas the particle $X$ is characterized by a narrow intrinsic width. Two events with 
a four-jet resonance mass $m_{4j}\approx 8$ TeV and an average dijet mass 
$\overline m_{2j}\approx 2$ TeV of the two dijet resonances were observed by the CMS 
\cite{CMS:2020zti} and reanalyzed in \cite{CMS:2025hpa}. They stand out with a local 
significance above 3.6 standard deviations owing to extremely small QCD background.
An event with $m_{4j}= 6.6$ (5.8) TeV and $\overline m_{2j}= 2.2$ (2.0) TeV was 
also reported by the ATLAS \cite{ATLAS:2023ssk} (CMS \cite{CMS:2020zti}). The wider spread 
of the measured $m_{4j}$ and the similar values of $\overline m_{2j}$ suggest that 
the mediators are potentially broad resonances \cite{CMS:2025hpa}. Another excess occurs 
in the non-resonant search for dijet resonances via $pp \to XX \to (jj)(jj)$; Figures~10 
and 13 in \cite{CMS:2020zti} indicate the excess for a four-jet mass between 3 and 4 TeV 
with an average dijet resonance mass 0.95 TeV. The reinterpretation in \cite{CMS:2025hpa} 
attributed it to a resonant production corresponding to $m_{4j}\approx 3.6$ TeV and 
$\overline m_{2j}\approx 1.0$ TeV with a local (global) significance of 3.9 (2.2) standard 
deviations. 

On the theoretical side, various new physics models have been proposed to explain the 
excesses of di-dijet events. For example, color-sextet diquark scalars $S_{uu}$, that 
decay into two vector-like quarks $\chi$, address the events with $m_{4j}\approx 8$ TeV and 
$\overline m_{2j}\approx 2$ TeV \cite{Dobrescu:2018psr,Dobrescu:2019nys,Dobrescu:2024mdl}. 
For the follow-up analyses on the channels $S_{uu} \to \chi\chi \to (Wb)(Wb)\to (jjb)(jjb)$ 
with six-jet final states and on the single vector-like quark channel 
$S_{uu} \to u\chi\to u(Wb,\; Zt,\; ht)$, refer to \cite{Duminica:2025lte,Costache:2025bjc} 
and \cite{Filip:2026rsw}, respectively, where $W$ ($b$, $Z$, $t$, $h$) denotes a $W$ boson 
($b$ quark, $Z$ boson, $t$ quark, Higgs boson). In terms of supersymmetry with $R$-parity 
violation, the 8 TeV (2 TeV) resonance is identified as a down-squark of the second or 
third generation (right-handed squark of the first generation) \cite{Bittar:2025rcw}. 
Pair-produced color-octet scalars $\Theta$ in $pp\to \Theta\Theta \to (q\bar q)(q\bar q)$ 
account for the dijet resonance with the mass 0.95 TeV \cite{Dobrescu:2025hyv}. Both the 
3.6 TeV and 0.95 TeV resonances were assigned to heavy gluons (color-octet vectors, i.e., 
colorons \cite{Hill:1993hs}) from a $SU(3)_1\times SU(3)_2\times SU(3)_3$ gauge group 
\cite{Crivellin:2022nms}, which is spontaneously broken to the $SU(3)_c$ group in the 
Standard Model (SM). The scenario with two color-sextet diquark scalars was also attempted 
in the same reference \cite{Crivellin:2022nms}. In summary, none of the models provides a 
comprehensive picture which accommodates all the involved resonance masses ranging from 1 
to 8 TeV. 

We have explored recently the phenomenological impacts of the sequential fourth generation 
model (SM4) with superheavy quarks. This model is motivated by the dynamical 
interpretation of the SM flavor structure \cite{Li:2023dqi,Li:2023yay,Li:2023ncg,Li:2024awx}; 
the mass hierarchy and the distinct mixing patters of quarks and leptons are dictated by the 
analyticity of SM dynamics. The SM4 is the most economical extension of the SM, to
which no additional free parameters need to be introduced; the mass $m_{t'}\approx 200$ TeV 
($m_{b'}= 2.7$ TeV) of a fourth-generation quark $t'$ ($b'$) was demanded by the 
dispersion relations for the mixing between the neutral quark states $t'\bar u$ and 
$\bar t' u$ ($b'\bar d$ and $\bar b' d$) \cite{Li:2023fim}. The mass $m_{\tau'}=270$ GeV 
($m_4=170$ GeV) of a fourth-generation charged (neutral) lepton $\tau'$ ($\nu'$) was 
predicted by examining the dispersion relation for the decay $\tau'\to \nu\bar t d$ 
($t\to d e^+\nu'$), $\nu$ ($e^+$) being a light neutrino (a positron) \cite{Li:2024xnl}.
Our observation echoes the ``$S$-matrix bootstrap conjecture" advocated by Geoffrey Chew in 
1960s \cite{Chew:1962mpd} that a well-defined infinite set of self-consistency conditions 
(based on analyticity, unitarity, causality, etc.) determines uniquely the aspects of 
particles in nature \cite{Cushing:1985zz,vanLeeuwen:2024uzj}. 

It has been shown that fermions with masses above a TeV scale form bound states in a Yukawa 
potential created by Higgs boson exchanges \cite{Hung:2009hy,Li:2025bon}. This is how 
superheavy fermions with large Yukawa couplings avoid the perturbative unitarity bound. A 
heavy scalar then appears as a composite of fourth-generation quarks, whose contributions 
to the Higgs boson production via gluon fusion and to the Higgs decay into a photon pair 
reduce to $O(10^{-3})$ and $O(10^{-2})$ of the top quark one \cite{Li:2023fim}, respectively. 
These estimates elucidated why superheavy fourth-generation quarks bypass the experimental 
constraints from Higgs boson production and decay 
\cite{Chen:2012wz,Eberhardt:2012gv,Djouadi:2012ae,Kuflik:2012ai}. Similar reasoning 
concludes that the SM4 also survives the experimental constraints from the oblique
parameters $S$, $T$ and $U$ \cite{Li:2024xnl} and from the parameter 
$\Delta\rho$, which is proportional to $T$ via the relation $\Delta\rho=\alpha T$, $\alpha$ 
being the fine structure constant. We postulate that the resonances in those 
LHC di-dijet events can be understood in terms of the bound states of fourth-generation 
quarks $b'$, including the ground, excited and multi-particle states. Our idea can be 
compared to the context of extra space-time dimensions with Kaluza-Klein excitations of 
gluons \cite{Crivellin:2022nms}, in which the dijet resonance is regarded as the lowest 
lying state, and the di-dijet resonance corresponds to a higher state. The difference is 
that a Yukawa potential allows only a finite number of bound states, while a Kaluza-Klein 
tower contains an infinite series of massive particles.

A $b'$ quark has a mass of 2.7 TeV at the electroweak scale, and a mass of 1.6 TeV
at the electroweak symmetry restoration scale of $O(10)$ TeV \cite{Li:2026gxw}, whose 
variation is governed by the two-loop renormalization-group (RG) evolution of the 
fourth-generation Yukawa couplings in the SM4 \cite{Hung:2009hy}. Hence, it is reasonable 
to assume the $b'$ quark mass $m_{b'}\approx 2.0$ TeV at a scale of $O(1)$ TeV. The 
production of four $b'$ quarks in $q\bar q$ annihilation is thus enhanced resonantly 
at a center-of-mass energy around 8 TeV. This $b'b'\bar b'\bar b'$ resonance is likely to  
be a tetraquark. A $b'\bar b'$ pair in the $b'b'\bar b'\bar b'$ system annihilates into two 
light jets via deep bound states under the Yukawa interaction. We evaluate 
the mass spectrum for $b'\bar b'$ bound states in the relativistic formalism 
\cite{Ikhdair:2012zz}, obtaining the ground state mass 0.94 TeV and the first excited state 
mass 2.1 TeV from the Yukawa potential characterized by the Higgs boson mass $m_H= 125$ GeV,
in consistency with the dijet resonance masses. The four-jet masses 6.6 TeV and 5.8 TeV 
might be associated with lower lying $b'b'\bar b'\bar b'$ tetraquarks, which can also decay 
into two $b'\bar b'$ excited states. Note that it is possible to interpret the four-jet 
event with the mass 5.8 TeV in the broad resonance hypothesis at a resonance mass around 8 TeV 
\cite{CMS:2025hpa}, whose width is not yet fully certain. The mass spectrum of the
$b'b'\bar b'\bar b'$ tetraquarks requests additional investigation, so the four-jet events 
with the masses 6.6 TeV and 5.8 TeV will not 
be the focus of the present work. The four-jet mass 3.6 TeV (between 3 to 4 TeV), far below 
the total mass of four $b'$ quarks, results from the non-resonant production of a 
$b'b'\bar b'\bar b'$ system, which decays only into two $b'\bar b'$ ground states. 


Our scenario represents a TeV-scale analogue of the detection of a fully charmed tetraquark 
$X(6900)$ in the four-muon channel at a GeV scale, $X(6900)\to J/\psi J\psi\to 4\mu$, by 
the LHCb Collaboration \cite{LHCb:2020bwg}, which was confirmed by the ATLAS and CMS 
Collaborations \cite{ATLAS:2023bft,CMS:2023owd} later. It was also 
identified via $X(6900)\to J/\psi \psi(2S)\to 4\mu$ \cite{ATLAS:2025nsd}, where one $J/\psi$ 
meson is replaced by an excited state $\psi(2S)$. A $X(6900)$ particle, appearing as a 
narrow resonance in the spectrum of the di-$J/\psi$ mass $M_{{\rm di}-J/\psi}= 6.9$ GeV in 
Fig.~2 of \cite{LHCb:2020bwg}, consists of four approximately on-shell charm quarks 
$cc\bar c\bar c$. It is analogous to the $b'b'\bar b'\bar b'$ resonance with the mass 8 TeV. 
The broad state centered at $M_{{\rm di}-J/\psi}= 6.6$ GeV in Fig.~2 of \cite{LHCb:2020bwg}, 
if interpreted as a lower lying state $X(6600)$ \cite{CMS:2023owd,CMS:2025fpt}, 
corresponds to the $b'b'\bar b'\bar b'$ resonances at 6.6 TeV and 5.8 TeV. If it is caused 
by a feed-down process \cite{LHCb:2020bwg}, the broad peak may mimic a non-resonant 
$b'b'\bar b'\bar b'$ production at 3.6 TeV. A vector $J/\psi$ with the mass $m_{J/\psi}=3.1$ 
GeV and an excited state $\psi(2S)$ with the mass $m_{\psi(2S)}= 3.7$ GeV parallel the ground 
and first excited states of a $b'\bar b'$ pair, respectively. Then the decay 
$X(6900)\to J/\psi \psi(2S)$ and the decay of the broad state into $J/\psi J/\psi$ are 
similar to the di-dijet events at 8 TeV and at 3.6 TeV with different dijet masses, 
respectively.

To verify our proposal, we need to simulate the signals from the process 
$(b'b'\bar b'\bar b')\to (b'\bar b')(b'\bar b')\to (jj)(jj)$ and the QCD background from 
multi-jet production. To avoid the complicated and tedious operation, we take an alternative 
approach, translating the SM4 to the ``effective" theories containing color-octet scalars 
and vectors available in the literature \cite{Dobrescu:2018psr,Crivellin:2022nms}. With the 
known masses and couplings in our SM4 setup, we will affirm that the relevant couplings from 
the matching between the full and effective theories can fit the LHC data well according to
\cite{Dobrescu:2018psr,Crivellin:2022nms}. The excesses of di-dijet events at 
various four-jet masses are then understood in the SM4 in this manner. The $b'$ quark 
systems are bound by an Yukawa interaction, instead of by QCD dynamics, so they can be 
color-octet states. The ground state labeled by $(n,l)=(1,0)$, $n$ ($l$) being the principal 
(angular momentum) quantum number, is either a pseudoscalar or a vector, where the latter is 
relevant to the current study. The first excited state with $(n,l)=(2,1)$ contains a $P$-wave 
scalar, which we will concentrate on. In line with our strategy, we assign a $b'b'\bar b'\bar b'$ 
resonance to a coloron of a mass 8 TeV, a $b'\bar b'$ excited state to a color-octet scalar 
of a mass 2 TeV \cite{Dobrescu:2018psr}, and a $b'\bar b'$ ground state to a coloron 
(color-octet vector) of a mass 1 TeV \cite{Crivellin:2022nms}. We remark that the di-quark 
model discussed in \cite{Dobrescu:2018psr,Dobrescu:2019nys,Dobrescu:2024mdl} is not favored 
from the viewpoint of the SM4, since the required $u\to b'$ transition in  $uu\to S_{uu}$ is 
highly suppressed by the $4\times 4$ CKM matrix element $V_{ub'}\sim 10^{-4}$ 
\cite{Li:2026gxw}.

The rest of the paper is organized as follows. We construct the $b'\bar b'$ mass spectrum 
in a relativistic formalism based on the Dirac equation in Sec.~\ref{sec2}, 
which yields the ground state (vector) mass 0.94 TeV and the first excited state (scalar) 
mass 2.1 TeV. The effective diagram involving colorons and color-octet scalars is matched 
to the Feynman diagram with four $b'$ quarks at a high energy scale in Sec.~\ref{sec3}. It is 
demonstrated that resultant effective couplings explain the di-dijet events around 
$m_{4j}\approx 8$ TeV. The similar analysis is applied to the matching of the effective 
diagram involving two different colorons to the same Feynman diagram with four $b'$ quarks
in Sec.~\ref{sec4}. It is corroborated that the excess around $m_{4j}\approx 3.6$ TeV can be
accounted for, if the mediator in the di-dijet events is treated as a virtual coloron 
of a mass 8 TeV considered in the previous case; namely, if these events are produced
non-resonantly. Section \ref{sec5} contains the summary.

\section{$b'\bar b'$ Mass Spectrum}
\label{sec2}

We deduce the mass spectrum for $b'\bar b'$ states bound by a Yuakawa potential 
\begin{eqnarray}
V(r)=-\alpha_Y\frac{e^{-m_H^{\ast} r}}{r},\label{yuk}
\end{eqnarray}
with the strength $\alpha_Y=m_{b'}^2/(4\pi v^2)$, the vacuum expectation value (VEV) $v=246$ 
GeV of a Higgs field, and the running Higgs boson mass $m_H^{\ast}$. The RG equations for 
fourth-generation Yukawa couplings have been presented in \cite{Hung:2009hy}, whose solutions 
imply a $b'$ quark mass 2.7 TeV at the electroweak scale $\mu\sim O(0.1)$ TeV, and 1.6 
TeV at the electroweak symmetry restoration scale $\mu\sim O(10)$ TeV \cite{Li:2026gxw}. 
The scale of $O(1)$ TeV we are interested in is located between the above two, so it 
is acceptable to take a $b'$ quark mass about 2.0 TeV. The quartic 
coupling $\lambda$ in the Higgs potential remains stable around its value 0.1 at the 
electroweak scale \cite{Hung:2009hy}, till it jumps to the fixed-point value 17 suddenly 
at a high scale. Therefore, $m_H^{\ast}=v\sqrt{2\lambda}\approx m_H$ ought to be an 
appropriate choice. It has been found \cite{Hung:2009hy} that heavy fermions, whose 
mass $m_Q$ meets the criterion $K_Q=m_Q^3/(4\pi v^2 m_H)>1.68$, form bound states. 
The above $b'$ quark mass satisfies the criterion $K_Q>1.68$ definitely, guaranteeing 
the existence of $b'\bar b'$ bound states. 


Properties of heavy quarkonium states, like $b'\bar b'$, have been explored intensively 
in the literature. It was shown \cite{Li:2023fim} that non-relativistic solutions 
\cite{Napsuciale:2021qtw} to the Schrodinger equation do not 
describe the $b'\bar b'$ bound states adequately owing to the Thomas collapse \cite{TH}. 
This inconsistency calls for a relativistic handling of the system 
\cite{Ikhdair:2012zz,Enkhbat:2011vp}. We employ Eq.~(28) in Ref.~\cite{Ikhdair:2012zz} for 
evaluating the eigenenergies $E_{n}$, which was derived from a Dirac equation with the 
potential in Eq.~(\ref{yuk}). The expression reads
\begin{eqnarray}
E_n&=&\frac{1}{\alpha_Y^2+4N^2}\Big\{\alpha_Y^2 W+4N^2S
+\sqrt{(\alpha_Y^2 W+4N^2S)^2-(\alpha_Y^2+4N^2)[(\alpha_Y W+2N^2m_H^\ast)^2+4N^2 MW]}\Big\},
\label{eb}
\end{eqnarray} 
with $N=n+1$, $W=C_s-M$ and $S=(C_s+\alpha_Y m_H^\ast)/2$, where $C_s$ is a parameter 
introduced by the spin symmetry of the Dirac equation \cite{Ginocchio:2005uv} and 
$M=m_{b'}/2$ is the reduced mass of the $b'\bar b'$ pair. The above solution holds under the 
conditions $M>E_n$ and $M+E_n>C_s$.

Note that the $b'\bar b'$ mass spectrum was attained for $C_s=0$ in \cite{Li:2023fim}, 
since the experimental information on these bound states was not yet taken into account. 
Here we intend to associate the resonances involved in the di-dijet production with the 
$b'\bar b'$ bound states; we thus tune $C_s$ to build the ground state mass $m_1\approx 1.0$ 
TeV for a color-octet vector, and then predict the first excited state mass $m_2$ for a 
color-octet scalar. This follows the standard practice in the literature 
(see, for instance, Ref.~\cite{Ahmadov:2021bwk}). The binding energy $E_n^b=E_{n}-M$ and 
the bound state mass $m_n=2m_{b'}+E_n^b$ are acquired from Eq.~(\ref{eb}) for an eigenenergy 
$E_{n}$. The value $C_s=-1.6 m_{b'}$ leads to the desired results $m_1= 0.94$ TeV and 
$m_2=2.1$ TeV. It is not unexpected that the binding energies of the 
$b'\bar b'$ bound states are of order of the constituent mass; the strength of the Yukawa 
interaction increases with constituent masses, so the order of the binding energies exhibits
a pattern different from that of QCD bound states governed by the gauge interaction.
For a reference, the $b'\bar b'$ state with $n=3$ gets a mass of 2.8 TeV.
A detection of four-jet events with dijet masses about 2.8 TeV in the future will 
be a direct support for our proposal. It is trivial to check that the above solutions satisfy 
the aforementioned conditions $M>E_n$ and $M+E_n>C_s$.


\section{Color-octet Scalars}

\label{sec3}

\begin{figure*}
\begin{center}
\includegraphics[scale=0.5]{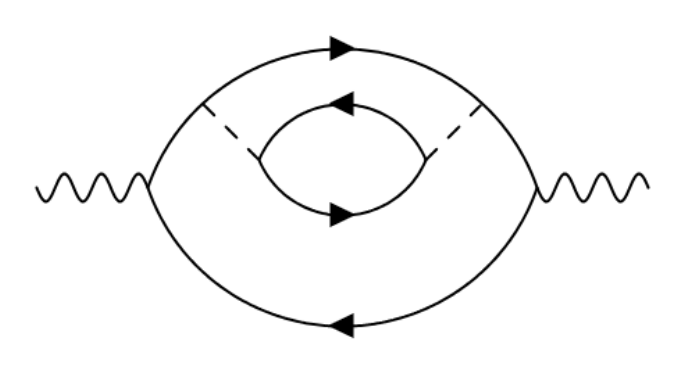}\hspace{1.0cm}
\includegraphics[scale=0.5]{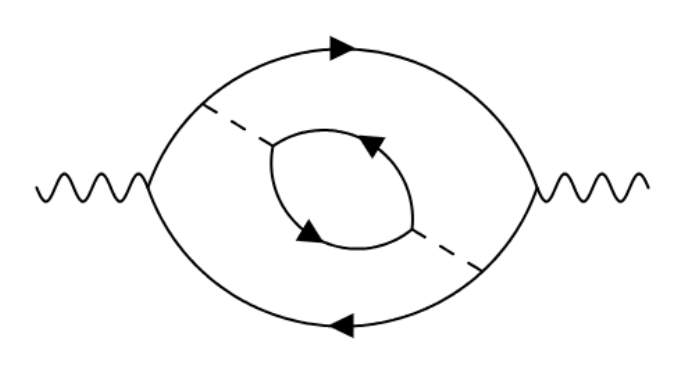}

(a) \hspace{6.0 cm} (b)
\caption{\label{fig1}
Feynman diagrams with internal fourth-generation quarks $b'$.}
\end{center}
\end{figure*}

We identify the diagrams, which contribute to the di-dijet processes at the LHC in the 
full and effective theories. A dominant leading-order Feynman diagram contains a virtual 
gluon from $q\bar q$ annihilation, that splits into a color-octet $b'\bar b'$ pair. 
A virtual Higgs boson is then emitted by one of the $b'$ quark lines, and splits into the 
second $b'\bar b'$ pair. If a gluon is substituted for the Higgs boson, the amplitude will 
be down by a power of $g_s^2/g_{b'}^2\ll 1$, $g_s$ ($g_{b'}$) being the strong coupling 
(the $b'$ quark Yukawa coupling). The corresponding effective diagram incorporates a 
coloron $Y$ as the mediator, which splits into two color-octet scalars $\Theta$ proposed 
in \cite{Dobrescu:2018psr}. The above two frameworks are supposed to provide the equivalent 
description of the $q\bar q\to q\bar q$ process at a high scale in the matching procedure, 
which is explicitly displayed in Figs.~\ref{fig1}(a) and \ref{fig1}(b) with intermediate 
$b'$ quarks in the full theory and in Fig.~\ref{fig2}(a) with $Y$ and $\Theta$ in the 
effective theory. The imaginary part of Fig.~\ref{fig2}(a) is directly related to the 
$Y\to \Theta\Theta$ cross section, which we concern. The contributions from 
Fig.~\ref{fig1}(a) and \ref{fig1}(b) can be evaluated without free parameters in the SM4, 
such that the effective coupling associated with $Y$ and $\Theta$ interaction can be 
extracted unambiguously from the matching. It will be elaborated that this effective 
coupling does accommodate the observed di-dijet excess at the four-jet mass 
$m_{4j}\approx 8$ TeV.

We start with the self-energy correction from $b'$ quarks to a Higgs boson propagator,
which is a sub-diagram of Figs.~\ref{fig1}(a) and \ref{fig1}(b). This one-loop integral 
is written as 
\begin{eqnarray}
i\Delta&=&-\left(-i\frac{g_{b'}}{\sqrt{2}}\right)^2{\rm Tr}(I_c)
\int\frac{d^4l}{(2\pi)^4}\frac{{\rm Tr}[i(\not l+m_{b'})i(\not l-\not p_H+m_{b'})]}
{(l^2-m_{b'}^2)[(l-p_H)^2-m_{b'}^2]}\nonumber\\
&\approx& \frac{3i}{8\pi^2}g_{b'}^2\Lambda^2,\label{del}
\end{eqnarray}
where the overall minus sign arises from the quark loop, ${\rm Tr}(I_c)$ denotes a trace
in color flow, $p_H$ is the Higgs boson momentum, and the ultraviolet cutoff 
$\Lambda\gg m_{b'}, m_H$ for the loop momentum does not exceed the symmetry restoration scale.  
Next we compute the self-energy correction from the dressed Higgs boson to a $b'$ quark 
propagator in Fig.~\ref{fig1}(a),
\begin{eqnarray}
i\Sigma(p_{b'})&=&\left(-i\frac{g_{b'}}{\sqrt{2}}\right)^2\int\frac{d^4l}{(2\pi)^4}
\frac{i(\not p_{b'}-\not l+m_{b'})}{(l-p_{b'})^2-m_{b'}^2}
\frac{i}{l^2-m_H^2}i\Delta\frac{i}{l^2-m_H^2}\nonumber\\
&\approx&-i\frac{g_{b'}^2}{32\pi^2}\Delta
\frac{\not p_{b'}+m_{b'}}{p_{b'}^2-m_{b'}^2}\ln\frac{m_{b'}^2}{m_H^2},\label{sig}
\end{eqnarray}
where $p_{b'}$ is the $b'$ quark momentum, and only the piece enhanced by the soft 
logarithm $\ln(m_{b'}^2/m_H^2)$ is kept. 

We then perform the final loop integration for Fig.~\ref{fig1}(a),
\begin{eqnarray}
\Pi_s(p)&=&-(-ig_s)^2{\rm Tr}(T^bT^a)\int\frac{d^4l}{(2\pi)^4}
{\rm Tr}\left[\frac{i(\not l-\not p+m_{b'})}{(l-p)^2-m_{b'}^2}
\gamma_\nu\frac{i(\not l+m_{b'})}{l^2-m_{b'}^2} i\Sigma(l)
\frac{i(\not l+m_{b'})}{l^2-m_{b'}^2}\gamma_\mu\right]\nonumber\\
&=&i\frac{g_s^2g_{b'}^2}{(16\pi^2)^2}\Delta 
\ln\frac{m_{b'}^2}{m_H^2}\delta^{ab}g_{\mu\nu}
\left\{\int_0^1dx(1-x)\left[\ln\frac{\Lambda^2}{m_{b'}^2-x(1-x)p^2}-\frac{3}{2}\right]
-\frac{1}{2}\right\},
\end{eqnarray}
with the virtual gluon momentum $p$. The combination with the self-energy correction 
to the lower $b'$ quark line leads to
\begin{eqnarray}
\Pi_s(p)&=&i\frac{g_s^2g_{b'}^2}{(16\pi^2)^2}\Delta 
\ln\frac{m_{b'}^2}{m_H^2}\delta^{ab}g_{\mu\nu}
\left\{\int_0^1dx\left[\ln\frac{\Lambda^2}{m_{b'}^2-x(1-x)p^2}-\frac{3}{2}\right]-1\right\}.
\label{ps1}
\end{eqnarray}

We come to the vertex correction from the dressed Higgs boson in Fig.~\ref{fig1}(b), 
\begin{eqnarray}
V(p_1,p_2)&=&\left(-i\frac{g_{b'}}{\sqrt{2}}\right)^2\int\frac{d^4l}{(2\pi)^4}
\frac{i(\not p_2-\not l+m_{b'})}{(l-p_2)^2-m_{b'}^2}
\gamma_\nu T^b\frac{i(\not p_1-\not l+m_{b'})}{(l-p_1)^2-m_{b'}^2}
\frac{i}{l^2-m_H^2}i\Delta\frac{i}{l^2-m_H^2}\label{v2}\\
&\approx&\frac{g_{b'}^2}{32\pi^2}\Delta\frac{(\not p_2+m_{b'})\gamma_\nu T^b(\not p_1+m_{b'})}
{(p_2^2- m_{b'}^2)(p_1^2- m_{b'}^2)}\ln\frac{m_{b'}^2}{m_H^2},\nonumber
\end{eqnarray}
where $p_1$ and $p_2$ are the momenta carried by the two $b'$ quarks with $p=p_1-p_2$,
and only the piece enhanced by the soft logarithm is retained. 
The final loop integration for Fig.~\ref{fig1}(b) yields
\begin{eqnarray}
\Pi_v(p)&=&-(-ig_s)^2\frac{g_{b'}^2}{32\pi^2}\Delta\ln\frac{m_{b'}^2}{m_H^2}
{\rm Tr}(T^bT^a)\int\frac{d^4l}{(2\pi)^4}
{\rm Tr}\left[\frac{i(\not l-\not p+m_{b'})}{(l-p)^2-m_{b'}^2}
\frac{(\not l-\not p+m_{b'})\gamma_\nu (\not l+m_{b'})}
{[(l-p)^2- m_{b'}^2](l^2- m_{b'}^2)}
\frac{i(\not l+m_{b'})\gamma_\mu}{l^2-m_{b'}^2}\right]
\nonumber\\
&=&-i\frac{g_s^2 g_{b'}^2}{(16\pi^2)^2}\Delta\ln\frac{m_{b'}^2}{m_H^2}\delta^{ab}g_{\mu\nu}
\int_0^1 dx\left[\ln\frac{\Lambda^2}{m_{b'}^2-x(1-x)p^2}-1\right]. \label{pv1}
\end{eqnarray}
It is noticed that the ultraviolet logarithms in Eqs.~(\ref{ps1}) and (\ref{pv1}) cancel each 
other as a consequence of the current conservation. 

We have the quark-level result $\Pi(p)=\Pi_s(p)+\Pi_v(p)$,
\begin{eqnarray}
\Pi(p)&=&-i\frac{3g_s^2g_{b'}^2}{2(16\pi^2)^2}\Delta 
\ln\frac{m_{b'}^2}{m_H^2}\delta^{ab}g_{\mu\nu}\nonumber\\
&=&-i\frac{9g_s^2g_{b'}^4}{(16\pi^2)^3}\ln\frac{m_{b'}^2}{m_H^2} 
\Lambda^2\delta^{ab}g_{\mu\nu},
\end{eqnarray}
where the expression of $\Delta$ in Eq.~(\ref{del}) has been inserted.
The $q\bar q\to q\bar q$ annihilation amplitude is then given, in the full theory, by
\begin{eqnarray}
A&=&(-ig_s)^2\bar q\gamma^\mu T^a q\frac{-i}{p^2}
(-i)\frac{9g_s^2g_{b'}^4}{(16\pi^2)^3}\ln\frac{m_{b'}^2}{m_H^2} \Lambda^2\delta^{ab}g_{\mu\nu}
\frac{-i}{p^2}\bar q\gamma^\nu T^b q\nonumber\\
&=&-i\frac{9g_s^4g_{b'}^4}{(4\pi)^6}\ln\frac{m_{b'}^2}{m_H^2}
\frac{\Lambda^2}{(p^2)^2}\bar q\gamma^\mu T^a q\bar q\gamma_\mu T^a q.\label{pst}
\end{eqnarray}



The same amplitude can be formulated in terms of colorons $Y$ with the mass $m_Y$ and 
color-octet scalars $\Theta$ with the mass $m_\Theta$ in the effective theory. The original 
framework starts with the spontaneously symmetry breaking of a $SU(3)_1\times SU(3)_2$ group 
down to the $SU(3)_c$ group by VEVs of colored scalar fields \cite{Bai:2010dj}. 
The mixing of the gauge fields in $SU(3)_1$ and $SU(3)_2$ defines a gluon field in the SM 
and a coloron field, whose mass is induced by the symmetry breaking. The ratio of the gauge 
couplings for the two gauge groups is parametrized as $\tan\theta$, $\theta$ being the mixing 
angle. It has been demonstrated that the di-dijet event number based on the cross section 
$\sigma(pp\to Y\to\Theta\Theta)$ inferred in the original framework amounts only up to 
$1.2\times 10^{-2}$ in 78 fb$^{-1}$ integrated luminosity \cite{Dobrescu:2018psr}, which is 
too low to account for the CMS data. We regard the effective coupling $g'$ involved in the
$q\bar q\to Y\to\Theta\Theta$ channel as an unknown to be fixed by the matching. This strategy 
is legitimate, for a $b'\bar b'$ excited state differs from the colored scalars in
\cite{Bai:2010dj} essentially; the former is not responsible for the generation of the coloron
mass, but a scalar, which is coupled to a coloron following the standard vector-scalar 
interaction. The outcome $g'>g_s$ will be delivered below. The event number in 
\cite{Dobrescu:2018psr}, which assumes $g'=g_s$, can then be scaled up by a factor 
$g'^4/g_s^4$ straightforwardly.

We work on the coloron-quark coupling and the coloron-scalar coupling 
\begin{eqnarray}
ig'\tan\theta,\;\;\;\;\frac{g'}{2}(\cot\theta-\tan\theta)f^{abc}(l_{1\mu}+l_{2\mu}),
\end{eqnarray}
respectively, where the index $a$ is associated with the coloron leg, and $b$ ($c$) is 
associated with the scalar leg with the outgoing (incoming) momentum $l_1$ ($l_2$). 
It has been observed that the cross section $\sigma(pp\to Y\to \Theta\Theta)$ is relatively 
insensitive to the angle $\theta$, staying around its maximum in the interval of 
$\tan\theta\in[0.25,0.45]$ (see Fig.~15 in \cite{Dobrescu:2018psr}), so $\tan\theta$ is set 
to $\tan\theta\approx 0.35$. In fact, the variation of $\tan\theta$ within the above range 
does not affect our estimation much. The scalar loop integral in Fig.~\ref{fig2}(a) reads
\begin{eqnarray}
\Pi_\Theta&=&\left(\frac{g'}{2}\right)^2(\cot\theta-\tan\theta)^2f^{acd}f^{bdc}
\int\frac{d^4l}{(2\pi)^4}\frac{(2l-p)_\mu (2l-p)_\nu}
{(l^2-m_{\Theta}^2)[(l-p)^2-m_{\Theta}^2]}
\nonumber\\
&\approx&-i\frac{g'^2N_c}{64\pi^2}(\cot\theta-\tan\theta)^2
\Lambda^2\delta^{ab}g_{\mu\nu},
\end{eqnarray}
with the coloron momentum $p$, where only the piece leading in powers of the ultraviolet 
cutoff $\Lambda$ is made explicit, and the identity $f^{acd}f^{bcd}=N_c\delta^{ab}$ has been 
applied.

The amplitude is then written, in the effective theory, as
\begin{eqnarray}
A&=&(ig'\tan\theta)^2\bar q\gamma^\mu T^a q\frac{-i}{p^2-m_Y^2}\Pi_\Theta\frac{-i}{p^2-m_Y^2}
\bar q\gamma^\nu T^b q\nonumber\\
&\approx&-i\frac{3g'^4}{64\pi^2}(1-\tan^2\theta)^2\frac{\Lambda^2}{(p^2)^2} 
\bar q\gamma^\mu T^a q\bar q\gamma_\mu T^a q,\label{t2}
\end{eqnarray}
at a high virtuality $p^2$. Comparing Eqs.~(\ref{pst}) and (\ref{t2}), we find their same 
dependencies on $\Lambda$ and $p^2$, and determine the effective coupling  
\begin{eqnarray}
g'=\frac{g_{b'}}{4\pi}\left[\frac{12\ln(m_{b'}^2/m_H^2)}
{(1-\tan^2\theta)^2}\right]^{1/4}g_s.
\end{eqnarray} 
The inputs $g_{b'}=11.5$ (corresponding to $m_{b'}=2.0$ TeV) and $m_H=125$ GeV generate
\begin{eqnarray}
g'\approx 2.8g_s.\label{gpa}
\end{eqnarray}
The perturbative expansion in terms of this effective coupling still makes sense,
since next-to-leading contributions are expected to be down by a factor of
$2.8^2\alpha_s/\pi\approx 0.2$ at a TeV scale.


Equation~(\ref{gpa}) indicates that the cross section $\sigma(pp\to Y\to \Theta\Theta)$ 
obtained in \cite{Dobrescu:2018psr} should be amplified by $2.8^4\approx 60$ times. 
In other words, the number of di-dijet events 
$1.2\times 10^{-2}\times 60\approx 0.7$ would be expected in 78 fb$^{-1}$ integrated 
luminosity and 1.3 in 138 fb$^{-1}$ integrated luminosity. Hence, the CMS di-dijet excess 
at $m_{4j}\approx 8$ TeV can be understood in our SM4 setup. We mention that the event 
number $1.2\times 10^{-2}$ quoted from \cite{Dobrescu:2018psr} is based on the CT14 set of 
parton distribution functions \cite{Dulat:2015mca}, and on the nearly 100\% branching 
fraction for a $\Theta$ scalar decay into two gluons. Though the simulation was done 
for the scalar mass $m_\Theta=1.8$ TeV, the exact value of $m_\Theta$ is not crucial for 
the deduction of the effective coupling $g'$ as shown above.


\section{Color-octet Vectors}
\label{sec4}

\begin{figure*}
\begin{center}
\includegraphics[scale=0.5]{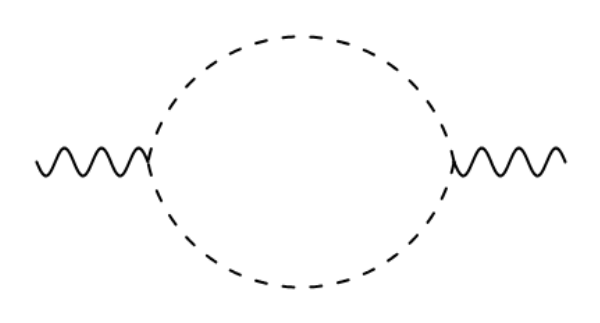}\hspace{1.0cm}
\includegraphics[scale=0.5]{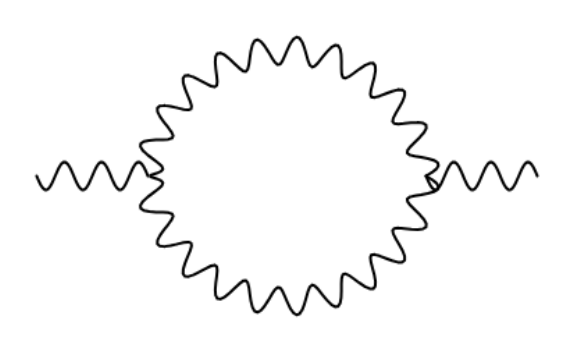}\hspace{1.0cm}

(a)\hspace{5.5cm}(b)\hspace{1.0cm}\mbox{}
\caption{\label{fig2}
Effective diagrams with (a) colorons (wavy lines) and color-octet scalars (dashed lines), 
and (b) two different colorons.}
\end{center}
\end{figure*}

We have assigned the $n=2$ excited state of a $b'\bar b'$ pair in a Yukawa potential to 
a color-octet scalar, and the $n=1$ ground state to a color-octet vector. This section 
will address the contribution of the latter to the di-dijet production at the four-jet mass 
3.6 TeV. The $q\bar q\to q\bar q$ annihilation amplitude in the full theory 
has been calculated in the previous section. As to the amplitude in the effective theory, we 
first derive the relevant coupling in the $SU(3)_1\times SU(3)_2\times SU(3)_3$ group proposed 
in \cite{Crivellin:2022nms}. The mass matrix for the $SU(3)$ gauge fields $G_i^{\mu a}$, with 
$i=1$, 2, 3 and $a=1,\cdots,8$, takes the form in the interaction basis, 
\begin{eqnarray}
L_M&=&\frac{1}{2}\left({\begin{array}{c}
G_1^{\mu a}\\G_2^{\mu a}\\G_3^{\mu a}\\
\end{array} } \right)^T
\left({\begin{array}{ccc}
   v_{12}^2g_1^2 & v_{12}^2g_1g_2 & 0 \\
   v_{12}^2g_1g_2  & (v_{12}^2+v_{23}^2)g_2^2 & v_{23}^2g_2g_3\\
   0  & v_{23}^2g_2g_3 & v_{23}^2g_3^2\\
  \end{array} } \right)
  \left({\begin{array}{c}
G_1^{\mu a}\\G_2^{\mu a}\\G_3^{\mu a}\\
\end{array} } \right),\label{bas}
\end{eqnarray}
where $g_i$ is the gauge coupling associated with the group $SU(3)_i$, and the VEVs $v_{12}$ 
and $v_{23}$ cause the spontaneous symmetry breaking, giving rise to the masses of the 
distinct colorons $X$ and $Y$.


The diagonalization of the above mass matrix leads to the mass eigenstates $g_1^{\mu a}$, 
$g_2^{\mu a}$ and $g_3^{\mu a}$; $g_1^{\mu a}$ corresponds to a massless SM gluon, and 
$g_2^{\mu a}$ ($g_3^{\mu a}$) corresponds to a coloron $X$ ($Y$) with the mass $m_X=0.95$ 
TeV ($m_Y=3.6$ TeV). The values of the coupling constants $g_1\approx 1$, $g_2\approx 10$ 
and $g_3\approx 15$ have been fitted from the ATLAS and CMS data in \cite{Crivellin:2022nms}. 
Demanding the nonvanishing eigenvalues to equal $m_X=0.95$ TeV and $m_Y=3.6$ TeV, we get 
two sets of solutions for the VEVs, $v_{12}=5.25$ TeV and $v_{23}=2.33$ TeV, and
$v_{12}=4.18$ TeV and $v_{23}=2.93$ TeV. It will be found that the first solution serves
better the purpose of interpreting the di-dijet excess. The diagonalization also specifies 
the transformation between $G_i^{\mu a}$ and $g_i^{\mu a}$,
\begin{eqnarray}
& &\left({\begin{array}{c}
G_1^{\mu a}\\G_2^{\mu a}\\G_3^{\mu a}\\
\end{array} } \right)=U
  \left({\begin{array}{c}
g_1^{\mu a}\\g_2^{\mu a}\\g_3^{\mu a}\\
\end{array} } \right),\nonumber\\
& &U=\left({\begin{array}{ccc}
   0.993 & -0.0946 & 0.0727 \\
   -0.0993 & -0.317 & 0.943\\
   0.0662 & 0.944 & 0.324\\
  \end{array} } \right),
\end{eqnarray}
where $U$ represents a $3\times 3$ rotation matrix with a unity determinant. 
Next we insert the above transformations into the products of 
field tensors $-(F_1^{\mu\nu}F_{1\mu\nu}+F_2^{\mu\nu}F_{2\mu\nu}+F_3^{\mu\nu}F_{3\mu\nu})/4$
and read off the coupling for the triple heavy-coloron vertex $YXX$,
\begin{eqnarray}
g_v=\sum_{i=1}^3 g_iU_{i2}^2U_{i3}=5.28.
\label{gvv}
\end{eqnarray}
The above result does not depend on which gauge field in $F_i^{\mu\nu}F_{i\mu\nu}$ 
is rotated under the transposed $U^T$, because of $U_{ij}=U^T_{ji}$ for a rotation matrix. We 
have ensured that the coupling $g_v$ is stable against the variation of $m_X$ ($m_Y$) around 
0.95 TeV (3.6 TeV).


The coloron loop integral in Fig.~\ref{fig2}(b) is expressed as
\begin{eqnarray}
\Pi_X&=&-g_v^2f^{acd}f^{bdc}
\int\frac{d^4l}{(2\pi)^4}\frac{(l+p)^\lambda g^{\mu\sigma}-(2p-l)^\sigma g^{\mu\lambda}-
(2l-p)^\mu g^{\sigma\lambda}}{l^2-m_X^2}
\frac{(l+p)_\lambda g_{\nu\sigma}-(2l-p)_\nu g^{\sigma\lambda}-
(2p-l)_\sigma g_{\lambda\nu}}{(l-p)^2-m_X^2}\nonumber\\
&=&-ig_v^2N_c\frac{9}{2}\frac{\Lambda^2}{16\pi^2}\delta^{ab}g_{\mu\nu},
\end{eqnarray}
with the $Y$ momentum $p$. Similarly, we pick up only the term leading in the ultraviolet 
cutoff $\Lambda$. The $q\bar q\to q\bar q$ amplitude is then given, in the effective 
theory, by
\begin{eqnarray}
A&=&(-ig'')^2\bar q\gamma^\mu T^a q\frac{-i}{p^2-m_Y^2}\Pi_X\frac{-i}{p^2-m_Y^2}
\bar q\gamma^\nu T^b q\nonumber\\
&\approx&-ig''^2g_v^2\frac{27}{32\pi^2}\frac{\Lambda^2}{(p^2)^2}
\bar q\gamma^\mu T^a q\bar q\gamma_\mu T^a q,\label{xa}
\end{eqnarray}
at a high virtuality $p^2$. Here we treat the coloron-quark coupling $g''$ as an unknown to 
be fixed by the matching.

The comparison between Eqs.~(\ref{xa}) and (\ref{pst}) designates the effective coupling
\begin{eqnarray}
g''=\frac{g_s^2g_{b'}^2}{8\pi^2 g_v}\sqrt{\frac{1}{6}\ln\frac{m_{b'}^2}{m_H^2}},
\end{eqnarray}
which takes the value, with the same inputs $g_{b'}=11.5$ and $m_H=125$ GeV,
\begin{eqnarray}
g''=0.31.
\label{g2}
\end{eqnarray}
Note that the branching fraction for the splitting $Y\to XX$ is almost 100\%, so the 
effective coupling extracted from the data in \cite{Crivellin:2022nms}, 
$g''=0.07/\sqrt{{\rm Br}(Y\to XX)}\approx 0.07$, is much smaller than Eq.~(\ref{g2}) from 
the matching to the SM4. Another set of VEVs gives $g''=0.19$, which is still larger than 0.07.
We suspect that the discrepancy is rooted in regarding the mediator as a 
resonance, and that the SM4 would require a model with $Y$ resonances of a mass 3.6 TeV to
generate too many di-dijet events. If the mediator is the same coloron of the mass 8 GeV 
considered in the previous section, the effective coupling in Eq.~(\ref{g2}) will not
change (if the parameters involved in Eq.~(\ref{bas}) are fixed), since the matching is 
performed at a high virtuality $p^2\gg m_Y^2$. The revised model then contains an off-shell 
mediator at the center-of-mass energy around 3.6 TeV, whose propagator introduces a suppression 
factor $3.6^2/8^2\approx 0.2$ as a naive estimate. This factor compensates the 
enhancement factor from the effective coupling, $0.31/0.07\approx 4.4$, ending up with 
$0.2\times 4.4 \approx 0.9$ close to unity. The resultant $pp\to Y^*\to XX$ cross section 
with $m_Y=8$ TeV is thus roughly equal to the one obtained in \cite{Crivellin:2022nms}, 
motivating us to postulate that the CMS di-dijet excess at $m_{4j}\approx 3.6$ TeV originates 
from a non-resonant production.


\section{SUMMARY}
\label{sec5}

We have explored the excesses of di-dijet events observed at the LHC, and attributed them
to the signals of the tetraquarks formed by fourth-generation quarks $b'$ with the mass 
$m_{b'}\approx 2$ TeV at few-TeV scales. The excesses at the four-jet masses $m_{4j}\approx 8$ 
TeV and 3.6 TeV can be explained in the same framework based on our SM4 setup; 
both are from the $b'b'\bar b'\bar b'$ production but through the resonant channel in the former 
and the non-resonant channel in the latter. The former (latter) then decays into color-octet 
scalars of a mass 2.0 TeV (color-octet vectors of a mass 0.95 TeV), which are associated with 
the first excited (ground) state of a $b'\bar b'$ pair. The above $b'\bar b'$ bound state 
masses have been verified by solving for the spectrum of a $b'\bar b'$ system in a Yukawa 
potential relativistically. The effective couplings involved in the events at various four-jet 
masses have been determined from the matching to the same Feynman diagrams with four 
intermediate $b'$ quarks. It is nontrivial that the distinct mechanisms responsible for 
resonant and non-resonant productions can be realized in our formalism simultaneously. 
The comprehensive picture for all the anomalous di-dijet 
events reported so far hints some underlying connection among them. We have pointed out that 
the di-quark proposal is not favored by the SM4 owing to the strong suppression on the 
$u\to b'$ transition by the CKM matrix element $V_{ub'}$, and that our scenario can be viewed 
as a TeV-scale version of the detection of a $X(6900)$ tetraquark in the four-muon modes 
$X(6900)\to (c\bar c)(c\bar c)\to 4\mu$ at a GeV scale.

The mass spectrum for a cluster of four superheavy fermions
bound by the Yukawa interaction deserves a separate project. This study will clarify
whether the excesses at $m_{4j}=6.6$ TeV and 5.8 TeV with the dijet 
mass about 2 TeV come from the lower lying $b'b'\bar b'\bar b'$ tetraquarks.
A $b'b'\bar b'\bar b'$ resonance can decay into two light jets, resulting in the process
$pp\to Y\to jj$. A $b'\bar b'$ bound state can be produced in $q\bar q$ 
annihilation, which also makes a dijet final state. The resonant dijet searches have been 
conducted at the LHC \cite{ATLAS:2017eqx,CMS:2018mgb}, and the ATLAS measurements reveal 
two events with dijet masses of approximately 8.0 and 8.1 TeV \cite{ATLAS:2017eqx}, near 
those in the di-dijet channels observed by the CMS Collaboration. There 
appears a weaker limit than expected in resonant dijet searches by the ATLAS 
\cite{ATLAS:2018qto} in a mass region slightly below 1 TeV. It was suggested 
\cite{Crivellin:2022nms} that this excess is due to the direct production of the same  
resonance $X$ with the mass 0.95 TeV in the collision $pp\to X\to jj$. We will analyze 
the above high-mass dijet events in the future.

\section*{Acknowledgement}

We thank T. Flacke for an inspiring discussion, which stimulated this project.
We also thank K.F. Chen, Y.J. Lin, C.T. Lu, S.M. Wang and R.L. Zhu for useful exchanges.
This work was supported in part by National Science and Technology Council of the Republic
of China under Grant No. NSTC-113-2112-M-001-024-MY3.


\end{document}